\documentclass[sigconf]{acmart}

\usepackage{subfigure}

\AtBeginDocument{%
  \providecommand\BibTeX{{%
    \normalfont B\kern-0.5em{\scshape i\kern-0.25em b}\kern-0.8em\TeX}}}

\copyrightyear{2020} 
\acmYear{2020} 
\setcopyright{acmcopyright}\acmConference[CIKM '20]{Proceedings of the 29th ACM International Conference on Information and Knowledge Management}{October 19--23, 2020}{Virtual Event, Ireland}
\acmBooktitle{Proceedings of the 29th ACM International Conference on Information and Knowledge Management (CIKM '20), October 19--23, 2020, Virtual Event, Ireland}
\acmPrice{15.00}
\acmDOI{10.1145/3340531.3412734}
\acmISBN{978-1-4503-6859-9/20/10}

\begin{document}
\fancyhead{}

%%
%% The "title" command has an optional parameter,
%% allowing the author to define a "short title" to be used in page headers.
\title{Personalized Bundle Recommendation in Online Games}

%%
%% The "author" command and its associated commands are used to define
%% the authors and their affiliations.
%% Of note is the shared affiliation of the first two authors, and the
%% "authornote" and "authornotemark" commands
%% used to denote shared contribution to the research.

\author{Qilin Deng}
\affiliation{%
  \institution{Fuxi AI Lab, NetEase Games}
  \streetaddress{Binjiang District}
  \city{Hangzhou}
  \state{Zhejiang}
  \country{China}}
\email{dengqilin@corp.netease.com}

\author{Kai Wang}
\affiliation{%
  \institution{Fuxi AI Lab, NetEase Games}
  \streetaddress{Binjiang District}
  \city{Hangzhou}
  \state{Zhejiang}
  \country{China}}
\email{wangkai02@corp.netease.com}

\author{Minghao Zhao}
\affiliation{%
  \institution{Fuxi AI Lab, NetEase Games}
  \streetaddress{Binjiang District}
  \city{Hangzhou}
  \state{Zhejiang}
  \country{China}}
\email{zhaominghao@corp.netease.com}

\author{Zhene Zou}
\affiliation{%
  \institution{Fuxi AI Lab, NetEase Games}
  \streetaddress{Binjiang District}
  \city{Hangzhou}
  \state{Zhejiang}
  \country{China}}
\email{zouzhene@corp.netease.com}

\author{Runze Wu}
\affiliation{%
  \institution{Fuxi AI Lab, NetEase Games}
  \streetaddress{Binjiang District}
  \city{Hangzhou}
  \state{Zhejiang}
  \country{China}}
\email{wurunze1@corp.netease.com}

\author{Jianrong Tao}
\affiliation{%
  \institution{Fuxi AI Lab, NetEase Games}
  \streetaddress{Binjiang District}
  \city{Hangzhou}
  \state{Zhejiang}
  \country{China}}
\email{hztaojianrong@corp.netease.com}

\author{Changjie Fan}
\affiliation{%
  \institution{Fuxi AI Lab, NetEase Games}
  \streetaddress{Binjiang District}
  \city{Hangzhou}
  \state{Zhejiang}
  \country{China}}
\email{fanchangjie@corp.netease.com}

\author{Liang Chen}
\affiliation{%
  \institution{Sun Yat-Sen University}
  \city{Guangzhou}
  \state{Guangdong}
  \country{China}}
\email{chenliang6@mail.sysu.edu.cn}

%%
%% By default, the full list of authors will be used in the page
%% headers. Often, this list is too long, and will overlap
%% other information printed in the page headers. This command allows
%% the author to define a more concise list
%% of authors' names for this purpose.
\renewcommand{\shortauthors}{Deng and Wang, et al.}

%%
%% The abstract is a short summary of the work to be presented in the
%% article.
\begin{abstract}
  In business domains, \textit{bundling} is one of the most important marketing strategies to conduct product promotions, which is commonly used in online e-commerce and offline retailers. Existing recommender systems mostly focus on recommending individual items that users may be interested in. In this paper, we target at a practical but less explored recommendation problem named bundle recommendation, which aims to offer a combination of items to users. To tackle this specific recommendation problem in the context of the \emph{virtual mall} in online games, we formalize it as a link prediction problem on a user-item-bundle tripartite graph constructed from the historical interactions, and solve it with a neural network model that can learn directly on the graph-structure data. Extensive experiments on three public datasets and one industrial game dataset demonstrate the effectiveness of the proposed method. Further, the bundle recommendation model has been deployed in production for more than one year in a popular online game developed by Netease Games, and the launch of the model yields more than 60\% improvement on conversion rate of bundles, and a relative improvement of more than 15\% on gross merchandise volume (GMV).
\end{abstract}

%%
%% The code below is generated by the tool at http://dl.acm.org/ccs.cfm.
%% Please copy and paste the code instead of the example below.
%%

\begin{CCSXML}
<ccs2012>
   <concept>
       <concept_id>10002951.10003317.10003338.10003343</concept_id>
       <concept_desc>Information systems~Learning to rank</concept_desc>
       <concept_significance>500</concept_significance>
       </concept>
   <concept>
       <concept_id>10002951.10003317.10003347.10003350</concept_id>
       <concept_desc>Information systems~Recommender systems</concept_desc>
       <concept_significance>500</concept_significance>
       </concept>
</ccs2012>
\end{CCSXML}

\ccsdesc[500]{Information systems~Learning to rank}
\ccsdesc[500]{Information systems~Recommender systems}

%%
%% Keywords. The author(s) should pick words that accurately describe
%% the work being presented. Separate the keywords with commas.
\keywords{recommender system; bundle recommendation; neural networks; deep learning; graph neural networks; link prediction}

%%
%% This command processes the author and affiliation and title
%% information and builds the first part of the formatted document.
\maketitle

\section{Introduction} \label{sec:introduction}

% background

Recommender system, which is an effective tool to alleviate the information overload, is widely used in modern e-commerce websites and online service business, e.g., Amazon, Taobao, Netflix. The basic goal of a recommender system is to find potentially interesting items for a user. Existing recommender systems mostly focus on recommending individual items to users, such as the extensive efforts on collaborative filtering that directly models the interaction between users and items. 

\begin{figure}[h]
  \centering
  \includegraphics[width=0.65\linewidth]{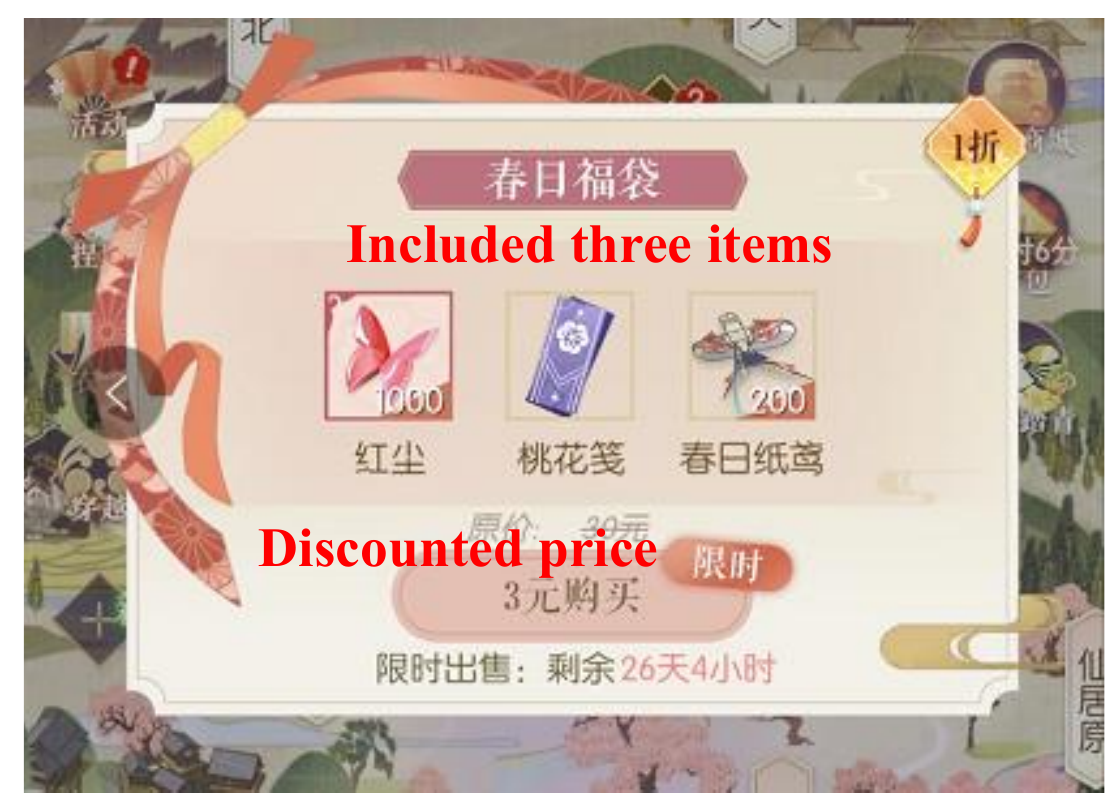}
  \caption{The Bundle GUI in the Game \emph{Love is Justice}.}
  \label{fig:gui}
  \Description{The GUI of bundle recommendation system.}
\end{figure}

In addition to consuming items individually, bundles are also ubiquitous in real-world scenarios. A bundle is a collection of items (products or services) consumed as a whole, and it usually reflects the frequent items which are appealing to most customers. In traditional business domains, e.g., supermarkets and offline retailers, it often takes bundling as a critical marketing strategy to attract customers and increase sales revenue. Moreover, the combination of items are especially ubiquitous on online service platforms, e.g., the music playlist on \emph{Spotify}, the book lists on \emph{GoodReads}, the boards of pins on \emph{Pinterest}, and the game bundles on \emph{Steam}.

% motivation

In Figure~\ref{fig:gui}, we show the bundle recommendation scenario in the online game \emph{Love is Justice}\footnote{https://yujian.163.com/}, a popular mobile game developed by \emph{Netease Games}\footnote{http://game.163.com/}, where users could impersonate specific roles and experience beautiful antique scenes and romantic plots, and their purchase and behavior data are extensively tracked by the game server. Here, we give some brief introductions to the Graphical User Interface (GUI) involved in this study. Once the user enters the homepage of the game interface (shown in the figure), the system will occasionally pop up a personalized discount bundle to attract the user, or the user himself can check the discount bundle when he wants. In addition, the items included in the bundle can also be purchased separately, although they are not discounted. According to our analysis of purchase statistics, more than \textbf{65\%} of game revenue comes from these discounted bundles, which also shows that it is profitable to increase the conversion rate of these personalized bundles.

In this paper, we address the problem of bundle recommendation in the context of online games, which aims to provide game players with the pre-defined bundles (combination of items) they are most likely to be interested in. Intuitively, this particular recommendation problem can be solved by treating bundles as "items" and then using traditional recommendation algorithms such as collaborative filtering. However, such straightforward solutions do not work well to capture user preference over bundles due to the following three difficulties:

\begin{itemize}
    \item \textbf{Data sparsity and cold-start.} Compared with user-item interactions, user-bundle interactions are usually more sparse due to the exponential combination characteristics of bundles and limited exposure resources. And only if the user is satisfied with the item combination or the discounted price is attractive, the user will have a strong willingness to buy the bundles rather than individual items, which makes the user-bundle interaction data appear more sparse.
    \item \textbf{Generalization over bundles.} Previous item recommendation algorithms may rely on item-level content features (e.g., category and brand in e-commerce), and user-item collaborative relationships. However, there is usually no informative bundle-level content features in the bundle recommendation scenario. This makes it difficult to provide the model's generalization ability in bundle preference prediction through a content-based model.
    \item \textbf{Correlation within the bundles.} The items within the bundle are usually highly correlated and compatible. Compared to typical item recommendation, the bundle recommendation problem is more complex considering that the user-bundle preference is a nontrivial combination of user-item preference. And directly modeling the interaction effect between items remains largely unexplored in the field of recommender systems.
\end{itemize}

% proposed model

\begin{figure}[h]
  \centering
  \includegraphics[width=0.7\linewidth]{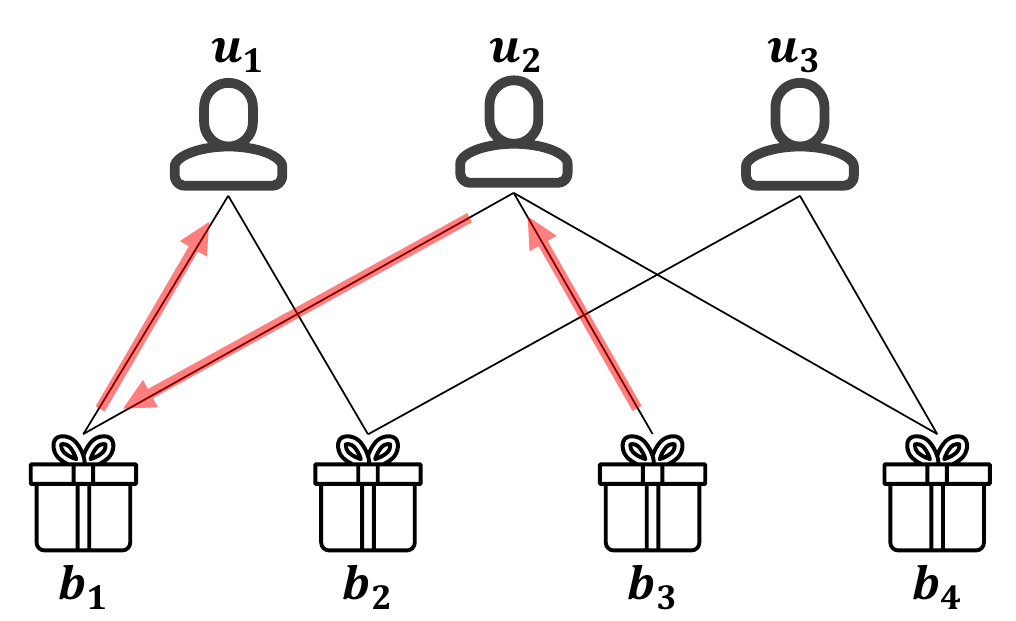}
  \caption{A toy example of a user-bundle bipartite graph with edges representing observed user-bundle interactions. The red arrow lines denote message passing paths.}
  \label{fig:example}
  \Description{The example graph.}
\end{figure}

Building on recent progress in deep learning on graph-structured data, we introduce a learning framework based on differentiable message passing on the user-item-bundle tripartite interaction graph constructed from historical data, and formalize the bundle recommendation problem as the link prediction problem in the tripartite graph. To alleviate the sparsity of user interactions on bundles, we integrate user-items interactions that provide additional information on user interests. To account for the compositional similarity between bundles, we derive the bundle representation by aggregating the item representations, which provides a natural good generalization ability over different bundles. We also model the correlation between bundle items in the form of learnable transformation parameters. Finally, we unify these improvements in our proposed framework named \emph{BundleNet}, and the multi-layer message passing structure can capture the \emph{high-order} and \emph{multipath} interactions over the user-item-bundle tripartite graph. As shown in Figure~\ref{fig:example}, bundle $b_3$ can reach user $u_1$ through the path $b_3 \to u_2 \to b_1 \to u_1$, and similar for bundle $b_4$. Moreover, compared with $b_3$, $b_4$ is a more reliable recommendation for $u_1$, since intuitively there is only one path existing between $u_1$ and $b_3$, while two paths connecting $u_1$ to $b_4$. Overall, the main contributions of this paper are summarized as follows:

\begin{itemize}
    \item We explore the promising yet challenging problem of bundle recommendation in the context of online games, and provide a practical case for the application of deep learning methods in the industry.
    \item We employ a differentiable message passing framework to effectively capture the user preferences for bundles, which can incorporate the intermediate role of items between users and bundles on the user-item-bundle tripartite graph.
    \item Extensive offline experiments on both in-game and other real-world datasets are conducted to verify the effectiveness of the proposed model. Further, we deploy the whole framework online and demonstrate its effective performance through online A/B Testing.
\end{itemize}

\section{Problem Definition} \label{sec:problem}

\begin{figure}[h]
  \centering
  \includegraphics[width=0.8\linewidth]{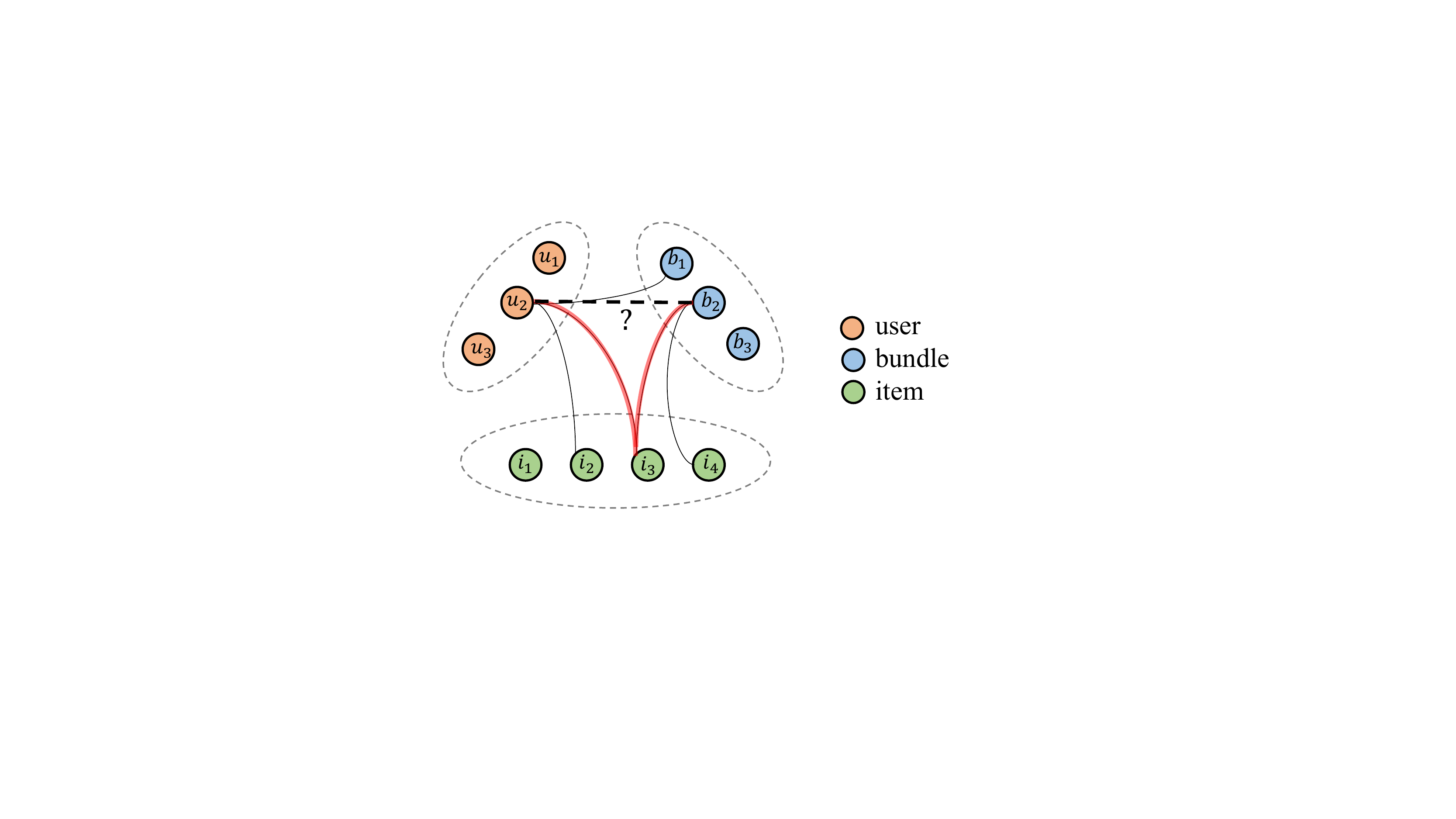}
  \caption{Problem Definition on the Tripartite Graph.}
  \label{fig:problem}
  \Description{problem.}
\end{figure}

Suppose we have users $\mathcal{U} = \{u_i|i=1,2,...,N_u\}$, items $\mathcal{I} = \{i_j|j=1,2,...,N_i\}$, and bundles $\mathcal{B} = \{b_k|k=1,2,...,N_b\}$, where the size of these sets is $|\mathcal{U}|=N_u, |\mathcal{I}|=N_i, |\mathcal{B}|=N_b$ respectively, and $N =N_u + N_i + N_b$. We also have the following three interaction graphs:

\begin{itemize}
    \item \textbf{User-Bundle Interaction}. A user can have an interaction (e.g., click, purchase) on an bundle, which is represented as a binary variable in the adjacency matrix $\mathbf{A}_{ub}$ of the user-bundle bipartite graph, i.e., 1 means the existence of the interaction and 0 otherwise. 
    \item \textbf{User-Item Interaction}. A user can also have an interaction (e.g., click, purchase) on a item, which is also represented as a binary variable in the adjacency matrix $\mathbf{A}_{ui}$ of the user-item bipartite graph, i.e., 1 means the existence of the interaction and 0 otherwise. 
    \item \textbf{Bundle-Item Interaction}. Each bundle $b$ is usually made up of several items, denoted as $b=\{i_1,i_2,,,,i_s\}$, where $s$ denotes the bundle size (larger than 1), and each item $i_i$ in the bundle belongs to the item set $\mathcal{I}$. We consider this containment relationship between bundles and their constituent items as the interaction relationship, just like the interaction between users and items. Similarly, the interaction is represented as a binary variable in the adjacency matrix $\mathbf{A}_{bi}$ of the bundle-item bipartite graph, i.e., 1 means the existence of the interaction and 0 otherwise. 
\end{itemize}

In the three interaction graphs discussed above, we actually have three bipartite graphs since there are only two types of entities in each interaction graph. Based on these bipartite graphs, in this paper, we introduce a user-item-bundle tripartite graph $\mathcal{G} = (\mathcal{U}, \mathcal{I}, \mathcal{B}, \mathcal{E})$, where $\mathcal{U}$ denotes user nodes, $\mathcal{I}$ denotes item nodes, $\mathcal{B}$ denotes bundle nodes and $\mathcal{E}$ is the corresponding link edges between these nodes, including user-item, user-bundle, bundle-item interactions, as illustrated in Figure~\ref{fig:problem}. 

The benefits that the introduction of item interaction information can bring us are twofold. On the one hand, a user's preference for a bundle could be inferred to a certain extent by his preference for items within the bundle, i.e., the preference over two bundles that share some items may be similar. On the other hand, the bundle contains the co-occurrence relationship between items, as items within a bundle are usually gathered based on a specific theme. If the item co-occurrence signal within the bundles can be properly utilized, we may learn a better recommendation model for individual items. These mutual correlations allow the performance of user-bundle and user-item recommendations to be mutually reinforced.

Based on the constructed tripartite graph $\mathcal{G}$, we define the bundle recommendation problem as a link prediction problem on graph $\mathcal{G}$. Essentially, this problem estimates the likelihood of an edge between a user node $u$ and a bundle node $b$ (e.g., the node $u_2$ and the node $b_2$ in Figure~\ref{fig:problem}), which represents how likely the user will be interested in the bundle. Formally, given the interaction graph $\mathcal{G}$, we propose a neural recommendation optimization model to learn an approximation function map $f$ as follows:

\begin{equation}
\hat{p} = f(u,b|\mathcal{G};\theta)
\end{equation}

Here, $\theta$ is the parameters of the neural model to be learned and $\hat{p}$ is the predicted likelihood that the user $u$ matches the bundle $b$, which will be specified in the following subsections.

\section{Methodology} \label{sec:proposed}

\begin{figure*}[htbp]
  \centering
  \includegraphics[width=0.8\textwidth]{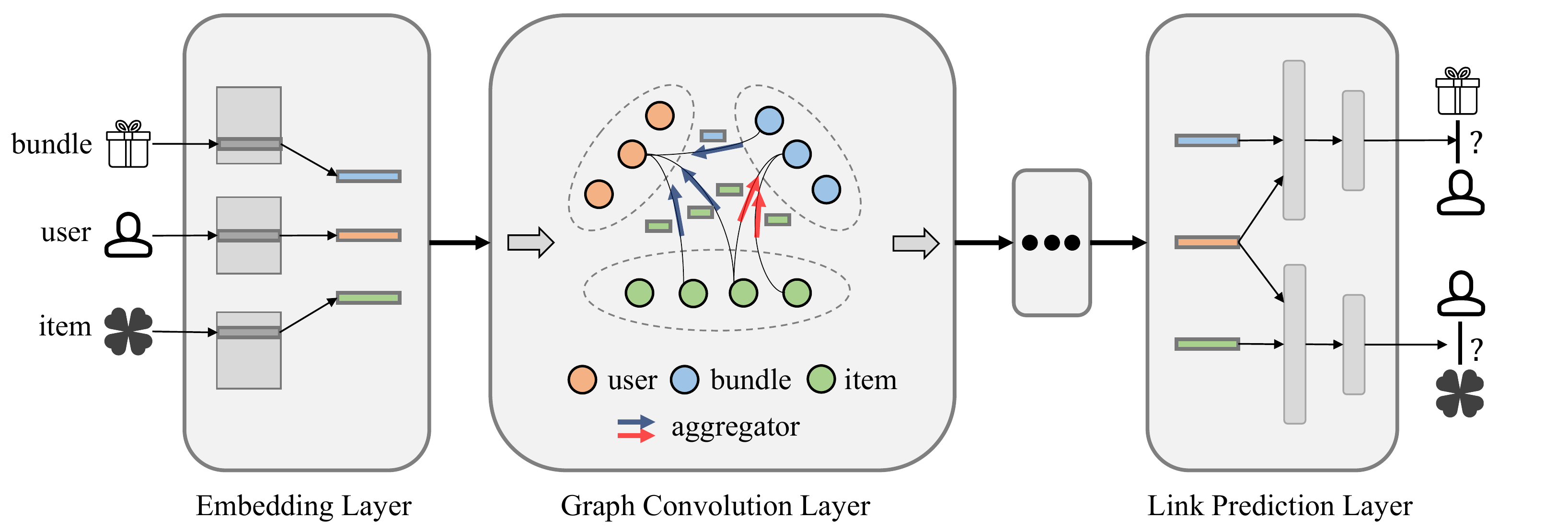}
  \caption{An illustration of the overall model framework of BundleNet.}
  \label{fig:framework}
  \Description{Illustrations of the BundleNet.}
\end{figure*}

We give the formal definition of the bundle recommendation problem above, in this section, we introduce the various components of the proposed model BundleNet in detail. The overall model framework is shown in Figure~\ref{fig:framework}.

\subsection{Embedding Layer}

Following existing research work \cite{cheng2016wide, he2017neural, wu2019neural}, for a tripartite graph $\mathcal{G} = (\mathcal{U}, \mathcal{I}, \mathcal{B}, \mathcal{E})$, we define $\mathbf{e}_{u} \in \mathbb{R}^{d}$, $\mathbf{e}_{i} \in \mathbb{R}^{d}$ and $\mathbf{e}_{b} \in \mathbb{R}^{d}$ as the embedding vectors of user node $u$, item node $i$ and bundle node $b$ respectively, where $d$ is the embedding size. It can be expressed as:

\begin{equation} \label{eq:embed}
\mathbf{e}_u = \mbox{EMBED} \left(u \right), \  \mathbf{e}_i = \mbox{EMBED} \left(i \right), \  \mathbf{e}_b = \mbox{EMBED} \left(b \right)
\end{equation}

Suppose we denote the one-hot feature vector for user $u$ as $x_u \in \mathbb{R}^{N}$, denote the embedding matrix of users as $\mathbf{E}_u \in \mathbb{R}^{N \times d}$, then we can obtain the user embedding vector of $u$ by $e_u=\mathbf{E}_u^T x_u$. Likewise, we can get the embedding representation of item nodes $\mathbf{e}_u$, bundle nodes $\mathbf{e}_b$, which is omitted here. We stack these node embeddings as the input representation for subsequent modules:

\begin{equation} \label{eq:input}
\mathbf{E} = [\mathbf{E}_u, \mathbf{E}_i, \mathbf{E}_b]
\end{equation}

\subsection{Graph Propagation Layer} \label{sec:gcn}

Inspired by recent convolutional neural networks that operate directly on graph-structured data, we use \emph{Graph Convolutional Networks} (GCNs) \cite{kipf2017semi} to process the tripartite graph data. GCN generalizes convolutions to graphs, which can naturally integrate both node attributes and topological structure in graphs, have been proved to be effective in representation learning for graph-structured data. Its propagation rule can be formulated as $\mathbf{Z}=f(\mathbf{X},\mathbf{A})$, where $\mathbf{X}$ denotes node feature matrix (node embedding in this work), $\mathbf{A}$ denotes adjacency matrix of the underlying graph structure, and $\mathbf{Z}$ denotes the encoded node representation. The single-layer propagation rule is:

\begin{equation} \label{eq:gcn}
\mathbf{Z}=f(\mathbf{X},\mathbf{A})=\sigma(\hat{\mathbf{A}} \mathbf{X} \mathbf{W})
\end{equation}

Here, $\hat{\mathbf{A}}={\tilde{\mathbf{D}}^{-1/2}} \tilde{\mathbf{A}} \tilde{\mathbf{D}}^{-1/2}$, with $\tilde{\mathbf{A}}=\mathbf{A}+\mathbf{I}$ and  $\tilde{D}_{ii}=\sum_j{\tilde{A}_{ij}}$, and $\hat{\mathbf{A}}$ can be calculated in a pre-processing step to speed up the model training. The $\sigma$ denotes an element-wise activation function such as the $\mbox{ReLU}(\cdot) = max(0, \cdot)$. In our case, the adjacency matrix of the user-item-bundle tripartite graph is constructed as follows: 

\begin{equation}
 \mathbf{A} = \left[
 {\begin{array}{*{20}{c}} 
 \mathbf{0} & \mathbf{A}_{ui} & \mathbf{A}_{ub} \\ 
 \mathbf{A}_{ui}^T & \mathbf{0} & \mathbf{A}_{bi}^T \\
 \mathbf{A}_{ub}^T & \mathbf{A}_{bi}  & \mathbf{0}
 \end{array}} \right] 
\end{equation}

where $\mathbf{A}_{ui}$, $\mathbf{A}_{ub}$ and $\mathbf{A}_{bi}$ denote the adjacency matrices of user-item, user-bundle and bundle-item interaction graph, respectively. Note that the elements on the main diagonal are all 0, since there is no self-loop connection edge. We can stack several layers to learn better hidden representations (high-order interactions) for graph nodes, with the following layer-wise propagation rule:

\begin{equation}
\mathbf{H}^{l+1}= \sigma(\mathbf{\hat{A}} \mathbf{H}^{l} \mathbf{W}^{l})
\end{equation}

where, $\mathbf{H}^{l}$ denotes input represeatation of graph nodes in the $l^{th}$ layer, $\mathbf{H}^{0}=\mathbf{E}$ is the embedding matrix given by formula~\ref{eq:input}, $\mathbf{W}^{l}$ denotes a layer-specific trainable weight matrix, $\mathbf{\hat{A}}$ is defined as above, $\sigma$ denotes an element-wise activation function such as the $\mbox{ReLU}(\cdot) = max(0, \cdot)$ and $\mathbf{H}^{l+1}$ is the output representation matrix in the $(l+1)^{th}$ layer. 

The standard GCN model is widely used in \emph{homogeneous} graph, however, the tripartite graph $\mathcal{G}$ is actually a \emph{heterogeneous} graph containing multiple types of nodes (user, item and bundle nodes) and multiple types of edges (user-item, user-bundle, bundle-item edges). Inspired by the \emph{Relational Graph Convolutional Network} (R-GCN) model \cite{schlichtkrull2018modeling}, we take the heterogeneous properties into account for our problem, and extends the GCN model to relational graphs, which could be considered as directed and labeled heterogeneous graphs. In our user-item-bundle tripartite graph setting, we consider three kinds of relations, i.e., the user-item interaction relation, the user-bundle interaction relation, and the bundle-item interaction relation, which consists of six relational edge\footnote{$\mathcal{R}$ contains relations both in canonical direction (e.g. user->item) and in inverse direction (e.g. item->user).}. The propagation rule for calculating the forward-pass update of an node $i$ in a relational graph is as follows:

\begin{equation} \label{eq:rgcn}
\mathbf{h}_{i}^{l+1}= \sigma \left( \mathbf{W}^{l} \mathbf{h}_i^{l} + \sum_{r\in \mathcal{R}} \sum_{j \in \mathcal{N}_i^r} \frac{1}{c_{i,r}} \mathbf{W}_{r}^{l} \mathbf{h}_{j}^{l} \right)
\end{equation}

where $\mathcal{N}_i^r$ denotes the set of neighbor indices of node $i$ under relation $r \in \mathcal{R}$, $\mathbf{W}^{l} \in \mathbb{R}^{d_0 \times d}$ denotes a trainable weight matrix. $c_{i,r}$ is a problem-specific normalization constant that can either be learned or chosen in advance (we use $c_{i,r}=|\mathcal{N}_i^r|$ in this work). 

\subsection{Link Prediction Layer}

After the iterative diffusion process propagated with $L$ layers, we obtain multiple representations for user node $u$, namely $\{\mathbf{h}_u^{1},...,\mathbf{h}_u^{L}\}$. The hidden representations obtained in different layers emphasize the messages passed over different connections or search depth in the graph, which makes them have different contributions in reflecting user preference. As such, we concatenate them to constitute the final user representations. Likewise, we can obtain the bundle and item representations by concatenating the bundle and item node representations $\{\mathbf{h}_b^{1},...,\mathbf{h}_b^{L}\}$, $\{\mathbf{h}_i^{1},...,\mathbf{h}_i^{L}\}$ learned by different layers. 

\begin{equation} \label{eq:hidden}
\mathbf{h}_u = \mathbf{h}_u^{1} || \cdots || \mathbf{h}_u^{L}, \quad
\mathbf{h}_i = \mathbf{h}_i^{1} || \cdots || \mathbf{h}_i^{L}, \quad
\mathbf{h}_b = \mathbf{h}_b^{1} || \cdots || \mathbf{h}_b^{L}
\end{equation}

where $||$ is the concatenation operation. In our experiments, we set $L=2$ since we found that stacking more than two convolutional layers did not improve performance.

The introduction of item information as a bridge role can make the model have a richer representation ability, which can be verified from the following experiments. Here, we simultaneously model user preferences for items and bundles, expecting their prediction performance to be mutually reinforced. Thus, with the final representations of users, items and bundles, we concatenate the latent vector representations of user $u$ and item $i$ as ${\mathbf{h}_u} || {\mathbf{h}_i}$, and feed them into a two-layer fully connected network (multilayer perceptron, MLP) to predict the preference $\hat{p}_{ui}$ of user $u$ to item $i$, and feed ${\mathbf{h}_u} || {\mathbf{h}_b}$ into another two-layer MLP to predict the preference $\hat{p}_{ub}$ of user $u$ to bundle $b$.

\begin{equation} \label{eq:pred2}
{\hat{p}_{ui}} = \mbox{sigmoid} \left( {\mathbf{W}_i^2} \ \mbox{ReLU} \left(  {\mathbf{W}_i^1}[{\mathbf{h}_u} || {\mathbf{h}_i}] + {\mathbf{b}_i^1} \right) + {\mathbf{b}_i^2} \right)
\end{equation}

\begin{equation} \label{eq:pred1}
{\hat{p}_{ub}} = \mbox{sigmoid} \left( {\mathbf{W}_b^2} \ \mbox{ReLU} \left(  {\mathbf{W}_b^1}[{\mathbf{h}_u} || {\mathbf{h}_b}] + {\mathbf{b}_b^1} \right) + {\mathbf{b}_b^2} \right)
\end{equation}

where $\mathbf{W}_i^1, \mathbf{W}_b^1 \in \mathbb{R}^{d_1 \times 2d_0}$ and $\mathbf{W}_i^2, \mathbf{W}_b^2 \in \mathbb{R}^{d_2 \times d_1}$ are corresponding weight matrices, $\mathbf{b}_i^1, \mathbf{b}_b^1 \in \mathbb{R}^{d_1}$ and $\mathbf{b}_i^2, \mathbf{b}_b^2 \in \mathbb{R}^{d_2}$ are corresponding bias, respectively. 

\subsection{Model Training}

\subsubsection{Loss Function} \label{sec:loss}

To train our BundleNet model, we adopt the Bayesian Personalized Ranking (BPR) \cite{rendle2009bpr} loss function. As a pairwise learning framework, BPR is an very pervasive personalized ranking criterion used in recommender systems and information retrieval community. It is based on the triplets data $\{u,p,n\}$, and the semantics is that user $u$ is assumed to prefer positive item $p$ over negative item $n$:

\begin{equation} \label{eq:bpr}
L_{BPR}(u,p,n) = -\ln{\sigma(\hat{p}_{up} - \hat{p}_{un})} + \lambda ||\Theta||_2^2
\end{equation}

where $\Theta$ denotes model parameters. $L2$ regularization is applied to prevent overfitting and $\lambda$ controls the regularization strength. 

\subsubsection{Multi-Task Learning}

By enforcing a common intermediate representation, Multi-Task Learning (MTL) can lead to better generalization and benefit all of the tasks, if the different problems are sufficiently related. This is obviously applicable in our scenario when we consider the user’s preferences for items and bundles at the same time. In our multi-task learning framework, we construct two kinds of triplets, i.e., user-item triplets $\{u, i^+, i^-\}$ and user-bundle triplets $\{u,b^+,b^-\}$, corresponding to two loss functions:

\begin{equation} \label{eq:loss}
L_1 = L_{BPR}(u, i^+, i^-), \quad L_2 = L_{BPR}(u, b^+, b^-)
\end{equation}

For triplets $\{u, i^+, i^-\}$, we first sample a user $u$, then sample the positive item $i^+$ from the bundles which $u$ have interaction history with, and a paired negative item $i^-$ from the rest of items. The similar process is performed for triplets $\{u, b^+, b^-\}$. In the experiments, we first use user-item interaction to minimize $L_1$ for pre-training, and then continue training with the bundle information to minimize $L2$ until convergence. An alternative strategy is to execute two gradient steps in turn to minimize $L_1$ and $L2$ \cite{chen2019matching}.

\subsubsection{Label Leakage Issue} \label{sec:leakage}

\begin{figure}[h]
\centering
\subfigure[Neighbor Aggregate in GCN.]{
\includegraphics[width=0.43\linewidth]{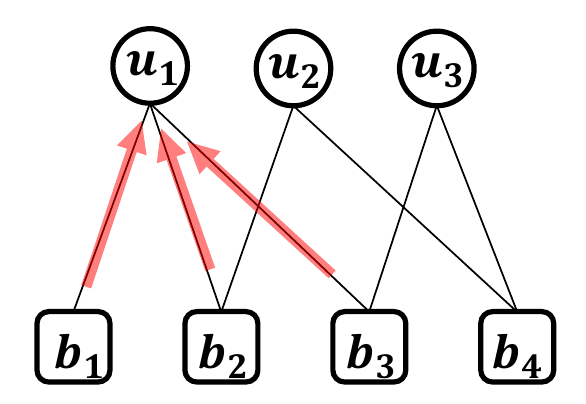}
\label{fig:leakage-1}
}
\quad
\subfigure[Mini-batch Setting for GCN.]{
\includegraphics[width=0.43\linewidth]{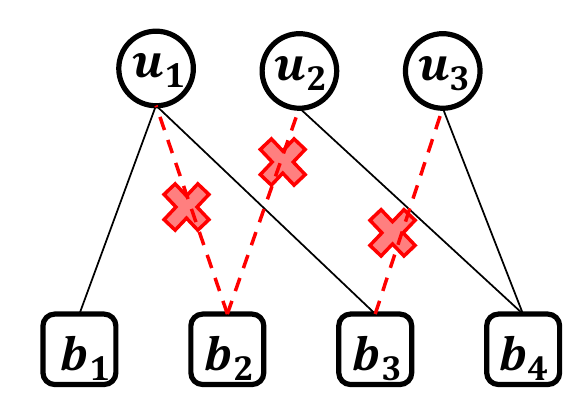}
\label{fig:leakage-2}
}
\caption{The Label Leakage Issue for Link Prediction.}
\end{figure}

We notice that the usual GCN-like model has a label leakage issue when it is used to solve the link prediction problem, which is also noted by \cite{zhang2019star}. Specifically, according to the design principle of the GCN, each node aggregate all neighbor information to update its self-representation. As shown in Figure~\ref{fig:leakage-1}, for example, when we want to predict an edge $e=(u_1, b_2)$, we have to learn the representation of both node $u_1$ and $b_2$. However, as the neighbor of $u_1$, we will aggregate the information of $b_2$ (along with $b_1$ and $b_3$) when we update the representation of node $u_1$. Similarly, when we update the representation of node $b_2$, we will also use the information of $u_1$. This means that the model actually tries to learn such a mapping $f_{\theta}(e,\cdots)=e$, leading to the label leakage issue, although it is a bit implicit in the GCN framework. The reason for this issue is that, when applied to link prediction problem, the usual GCN training method involves all elements of the entire graph to participate in training simultaneously (predict all existing edges in the graph, including $e$ of course). 

To avoid the label leakage issue, we need to make sure that the edge information (e.g., $e=(u_1, b_2)$) is not used when predicting the edge itself. Although the dropout technology can alleviate this, however, it does not essentially address the problem. Inspired by the training strategy in \cite{zhang2019star}, we adapt the usual (vanilla) \textbf{full-batch} training method of GCN to the \textbf{mini-batch} setting in the context of link prediction, following a \emph{sampling-deleting-predict} strategy. Instead of using all edges, at each training iteration step, we proceed as follows: first, we sample a batch of edges from the training graph (denoted as the red lines in the Figure~\ref{fig:leakage-2}), then we delete these sampled edges from the graph to ensure that they will not participate in the neighborhood aggregation operation during the training process. Finally, we perform the usual GCN training procedure on the modified graph, but only to predict those sampled (and deleted) edges, instead of all of the edges in the graph. With the mini-batch training method, we observe a substantial boost in link prediction performance, which can be observed in the compared results in following experiments. 

\subsection{Model Inference}

After the training is completed, in the inference stage, the test user id $u$, the bundle id $b$ and all item ids $\{i|i \in b\}$ within the bundle are fed into the model. Then, we can obtain the user's preference prediction for the bundle $\hat{p}_{ub}$ and for the included items $\{\hat{p}_{ui}| i \in b\}$. The final prediction of user’s preference for the bundle is:

\begin{equation} \label{eq:loss}
\hat{p} = \hat{p}_{ub} + \frac{1}{|b|}\sum_{i \in b} \hat{p}_{ui}
\end{equation}

For a newly released bundle, we could set $\hat{p}_{ub}$ to 0, and get the final preference prediction of the bundle just based on the user's prediction of the item, alleviating the cold start problem.

\section{Experiments} \label{sec:experiment}

\begin{table*}
  \centering
  \caption{Summary Statistics of the Datasets}
  \label{tab:datasets}
  \begin{tabular}{lcccccc}
    \toprule
    Datasets & \# users & \# bundles & \# items & \# user-bundle (density) & \# user-item (density) & \# bundle-item (density) \\
    \midrule  
    Steam    & 29,634  & 615     & 2,819    & 87,565 (0.48\%)   & 902,967 (1.08\%)    & 3,541 (0.20\%) \\
    Youshu   & 8,039   & 4,771   & 32,770   & 51,377 (0.13\%)   & 138,515 (0.05\%)    & 176,667 (0.11\%) \\
    NetEase  & 18,528  & 22,864  & 123,628  & 302,303 (0.07\%)  & 1,128,065 (0.05\%)  & 1,778,838 (0.06\%) \\
    Justice  & 25,470  & 234     & 278      & 117,873 (1.98\%)  & 379,384 (5.36\%)    & 483 (0.74\%) \\
    \bottomrule
  \end{tabular}
\end{table*}

\subsection{Datasets}

we evaluate all the models on three public datasets and one industrial dataset. 
The \emph{Steam} dataset is collected from the Steam\footnote{https://store.steampowered.com/} video game distribution platform by \cite{pathak2017generating}, where each bundle consists of several video games. The \emph{NetEase} dataset, provided by the work in \cite{cao2017embedding}, is crawled from the Netease Cloud Music\footnote{https://music.163.com/}, which enables users to construct the bundle (a list of songs) with a specific theme. The \emph{Youshu} dataset introduced by \cite{chen2019matching} is constructed by crawling data from a book review site Youshu\footnote{https://www.yousuu.com/}, where each bundle is a list of books constructed by website users. Finally, the \emph{Justice} dataset is collected from the mobile game \emph{Love is Justice} developed by Netease Games, where bundles are made up of props (virtual items) in the game. The statistics of datasets are briefly shown in Table~\ref{tab:datasets}.

\subsection{Baselines}

\begin{itemize}
  \item \textbf{BPR} \cite{rendle2009bpr}: This model is the basic pairwise ranking algorithm based on implicit feedback. We learn a BPR baseline model by user-bundle interactions, and optimize the BPR ranking loss under the matrix factorization framework. 
  \item \textbf{BundleBPR} \cite{velivckovic2017graph}: This is a bundle BPR model which makes use of the parameters learned through the item BPR.
  \item \textbf{DAM} \cite{chen2019matching}: This model is specially designed for the bundle recommendation, which jointly models user-bundle interactions and user-item interactions in a multi-task manner.
  \item \textbf{GCN-Bi} \cite{kipf2017semi}: The GCN model learned on the user-bundle bipartite graph with usual full-batch training method.
  \item \textbf{GCN-Bi-B} \cite{kipf2017semi}: The GCN-Bi model with our mini-batch training method introduced in section~\ref{sec:leakage}.
  \item \textbf{GCN-Tri}: The adapted GCN learned on our user-item-bundle tripartite graph with usual full-batch training method.
  \item \textbf{GCN-Tri-B}: The GCN-Tri mdoel with our mini-batch training method introduced in section~\ref{sec:leakage}.
  \item \textbf{BundleNet}: The adapted Relational GCN model learned on the user-item-bundle tripartite graph.
  \item \textbf{BundleNet-B}: The BundleNet model with our mini-batch training method introduced in section~\ref{sec:leakage}.
\end{itemize}

\subsection{Metrics}

We adopt the \emph{leave-one-out} evaluation in our experiments, which has been widely used in the literature \cite{rendle2009bpr, he2017neural}. For each user, we randomly hold-out one of her bundle interactions for testing and the remaining data for training. Since it is too time-consuming to rank all bundles for all users during the evaluation, we followed the common strategy \cite{he2017neural} that randomly samples 99 bundles that are not interacted by the user as negative samples, ranking the test bundle among the 99 bundles. To evaluate the recommendation performance of bundle models, we use several widely adopted ranking metrics in information retrieval: Recall, Mean Reciprocal Rank (MRR) and Normalized Discounted Cumulative Gain (NDCG). Specifically, the Recall@$K$ measures the number of positive bundles presenting within the top-$K$ recommendation list. The MRR@$K$ considers the rank position of the first positive bundle for the user within the top-$K$ recommendation list. And the NDCG@$K$ accounts for the position of the positive bundles by additionally considering the discount factor at lower ranks.

\subsection{Implementation}

All models were implemented in PyTorch\footnote{https://pytorch.org/} with the Adam optimizer \cite{kingma2014adam}. We also employ early stopping and dropout techniques to prevent over-fitting. The embedding size $d$ is fixed to $32$ for all models.  The hidden sizes $d_0, d_1, d_2$ are set to 64, 256, 128 respectively. The batch size for edge sampling is fixed to $1024$. We apply grid search for tuning the hyper-parameters of the models: the learning rate is tuned amongst \{$0.0001, 0.0005, 0.001, 0.005, 0.01$\}, the coefficient of $L2$ regularization is searched in \{$10^{-5}, 10^{-4},..., 1, 10^1$\}, and the dropout ratio in \{$0.0, 0.1,..., 0.5$\}. The set of possible hyper-parameter values was determined on early validation tests using subsets of the datasets that we then discarded from our analyses. 

\subsection{Results and Analysis}

\begin{table*}
  \small
  \centering
  \caption{Comparison of Results (for GCN-related models, including our model BundleNet, model names with and without the \textit{-B} suffix indicate that the mini-batch training method and the normal full-batch training method is used, respectively).}
  \label{tab:results}
  \begin{tabular}{lcccccccccccc}
    \toprule 
     & \multicolumn{3}{c|}{Steam} & \multicolumn{3}{c|}{Youshu} & \multicolumn{3}{c|}{NetEase} & \multicolumn{3}{c}{Justice} \\
    \cline{2-13}
     & Recall@5 & MRR@5 & NDCG@5 & Recall@5 & MRR@5 & NDCG@5 & Recall@5 & MRR@5 & NDCG@5 & Recall@5 & MRR@5 & NDCG@5 \\
    \midrule 
    BPR         & 0.9712 & 0.8002 & 0.8437 & 0.5409 & 0.3781 & 0.4278 & 0.3532 & 0.2086 & 0.2198 & 0.6735 & 0.4707 & 0.5223 \\
    BundleBPR   & \bf{0.9818} & \bf{0.8219} & \bf{0.8594} & 0.5912 & 0.3923 & 0.4408 & 0.4677 & 0.2765 & 0.3342 & 0.6925 & 0.5022 & 0.5482 \\
    DAM         & 0.9792 & 0.8016 & 0.8467 & \bf{0.5996} & \bf{0.4049} & 0.4532 & 0.4109 & 0.2424 & 0.2841 & 0.7117 & 0.4764 & 0.5349 \\
    GCN-Bi      & 0.9793 & 0.8069 & 0.8508 & 0.5753 & 0.3776 & 0.4267 & 0.3493 & 0.2037 & 0.2397 & 0.5578 & 0.3563 & 0.4061 \\
    GCN-Bi-B    & 0.9794 & 0.8106 & 0.8535 & 0.6001 & 0.4006 & 0.4503 & 0.4275 & 0.2597 & 0.3013 & 0.7427 & 0.4985 & 0.5594 \\
    GCN-Tri     & 0.9797 & 0.8012 & 0.8465 & 0.5893 & 0.3915 & 0.4408 & 0.3641 & 0.2138 & 0.2509 & 0.5718 & 0.3651 & 0.4172 \\
    GCN-Tri-B   & 0.9788 & 0.8092 & 0.8524 & 0.5924 & 0.3959 & \bf{0.4548} & \bf{0.5252} & \bf{0.3231} & \bf{0.3732} & \bf{0.7618} & \bf{0.5193} & \bf{0.5797} \\
    BundleNet   & 0.9788 & 0.8108 & 0.8536 & 0.5927 & 0.3962 & 0.4452 & 0.3579 & 0.2119 & 0.2481 & 0.5754 & 0.3742 & 0.4162 \\
    BundleNet-B & \bf{0.9848} & \bf{0.8859} & \bf{0.9112} & \bf{0.6241} & \bf{0.4247} & \bf{0.4668} & \bf{0.5142} & \bf{0.3114} & \bf{0.3616} & \bf{0.7705} & \bf{0.5545} & \bf{0.5807} \\
    \bottomrule 
  \end{tabular}
\end{table*}

We conduct extensive experiments on the datasets with the above benchmark methods to evaluate our model. We use $80\%$ of the data as training set to learn model parameters, $10\%$ as validation data to tune hyper-parameters and the rest $10\%$ as test set for final performance comparison. We repeat this procedure 10 times and report the average ranking values, which is summarized and shown in Table~\ref{tab:results}.  We can find that our proposed method outperforms the baseline methods significantly in all datasets. From the experimental result, we also have several interesting findings listed as follows:

\begin{itemize}
  \item The models of utilizing user-item interactions always outperform the models of not using this information, e.g., \emph{BundleBPR} is better than traditional \emph{BPR} and \emph{GCN-Tri} is better than \emph{GCN-Bi}. This result is obviously in line with our expectations and verifies the effectiveness of introducing item interaction in the bundle recommendation problem. This shows that leveraging the items as bridge signal/nodes to learn the representations of the users and/or bundles can alleviate the data sparsity problem.
  \item When considering modeling the bundle recommendation as a link prediction problem, models with mini-batch training method introduced in section~\ref{sec:leakage} always outperform the models without using this information, e.g., the \emph{GCN-Bi-B} and \emph{BundleNet-B} is better than \emph{GCN-Bi} and \emph{BundleNet}, respectively. We think the phenomenon is caused by the label leakage issue introduced above, and can be effectively alleviated through the mini-batch training trick. We believe that such comparison results bring us some useful inspirations, when using the GCN-like model for link prediction tasks.
  \item Our proposed model \emph{BundleNet} performs better than the state-of-the-art bundle recommendation method \emph{DAM}, which proves the effectiveness of modeling bundle recommendation as the link prediction problem in the user-item-bundle tripartite graph. Moreover, the \emph{BundleNet}/\emph{BundleNet-B} is slightly superior than the \emph{GCN-Tri}/\emph{GCN-Tri-B} for most datasets, which indicates that the heterogeneous characteristics of the user, item and bundle nodes and their interactions usually should not be ignored. However, in the \emph{NetEase} dataset, it is a bit worse. We guess that this is related to the characteristics of the data set, and it is worth further exploration.
\end{itemize}

\subsection{Ablation Study}

\begin{figure}[h]
  \centering
  \includegraphics[width=0.9\linewidth]{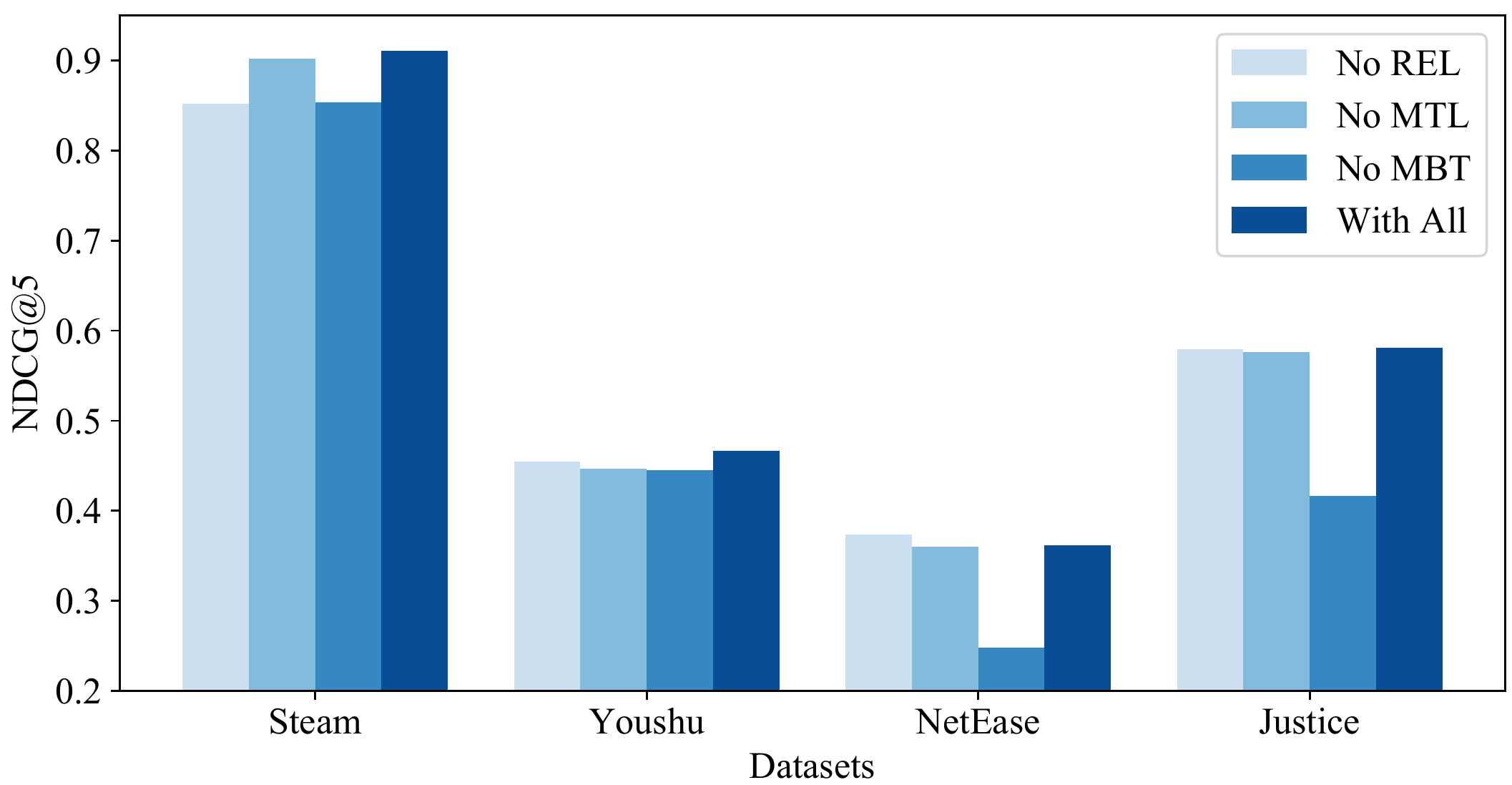}
  \caption{Performance Comparison of Major Designs.}
  \label{fig:ablation}
  \Description{Ablation Study.}
\end{figure}

In addition to the user-item-bundle tripartite graph, there are several designs involved in our model: the Relational GCN (\textbf{REL}) to account for heterogeneous properties of graph nodes, the multi-task learning (\textbf{MTL}) framework to model user’s preferences for items and bundles simultaneously, and the mini-batch training (\textbf{MBT}) method to solve the label leakage issue. To evaluate the effectiveness of these major designs, we carried out ablation studies as shown in Figure~\ref{fig:ablation}. The result demonstrates that these designs show different improvements for different datasets. For example, the \textbf{MBT} is crucial for NetEase and Justice, while both \textbf{REL} and \textbf{MBT} is beneficial to Steam. Meanwhile, Youshu is not very sensitive to these designs, which means its performance improvement mainly depends on the basic tripartite graph design.

\subsection{Online Deployment}

\begin{figure}[h]
  \centering
  \includegraphics[width=0.9\linewidth]{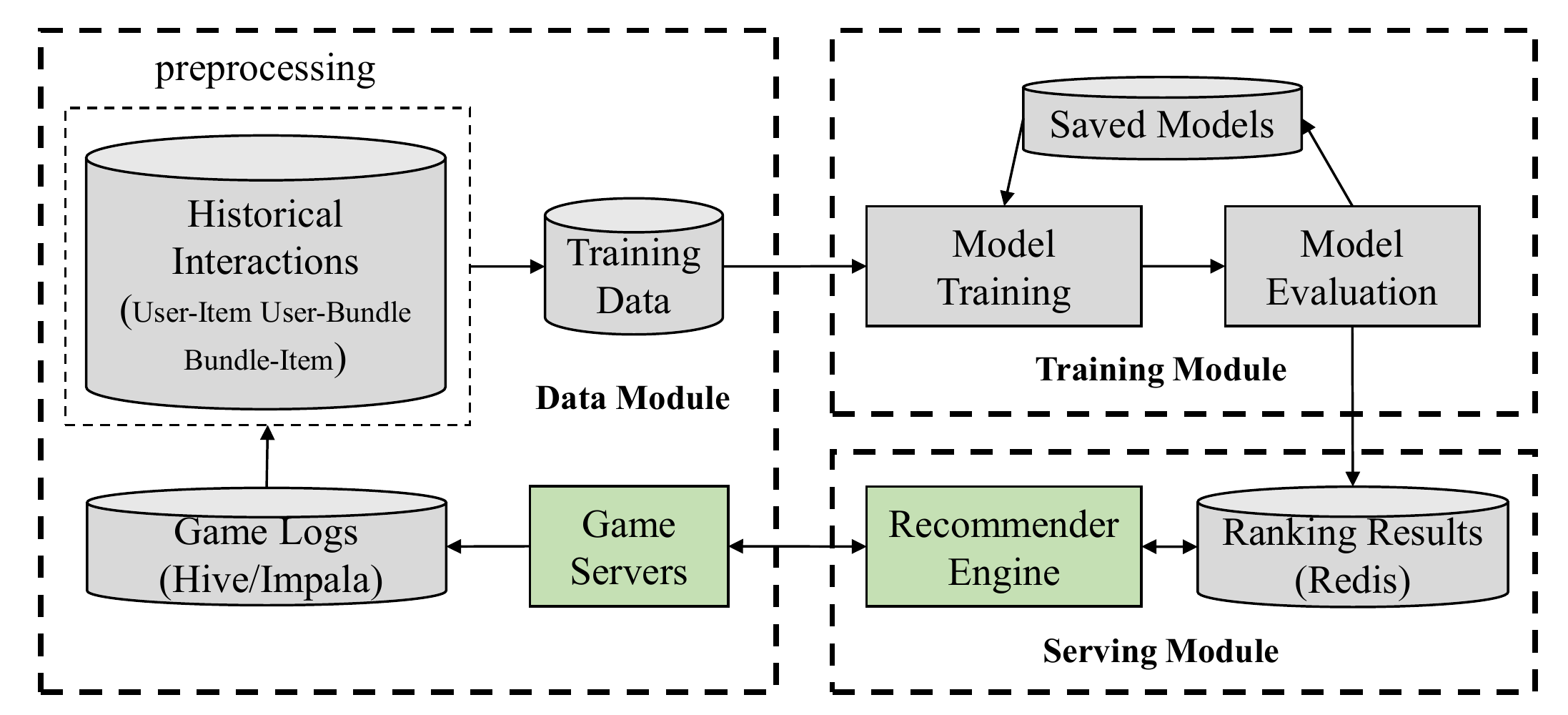}
  \caption{The Overview of Bundle Recommendation Workflow.}
  \label{fig:pipeline}
  \Description{Workflow.}
\end{figure}

\begin{figure*}[h]  
  \centering  
  \subfigure[Comparison of Conversion Rate (from left to right arranged by date).]{
  \includegraphics[width=0.48\textwidth]{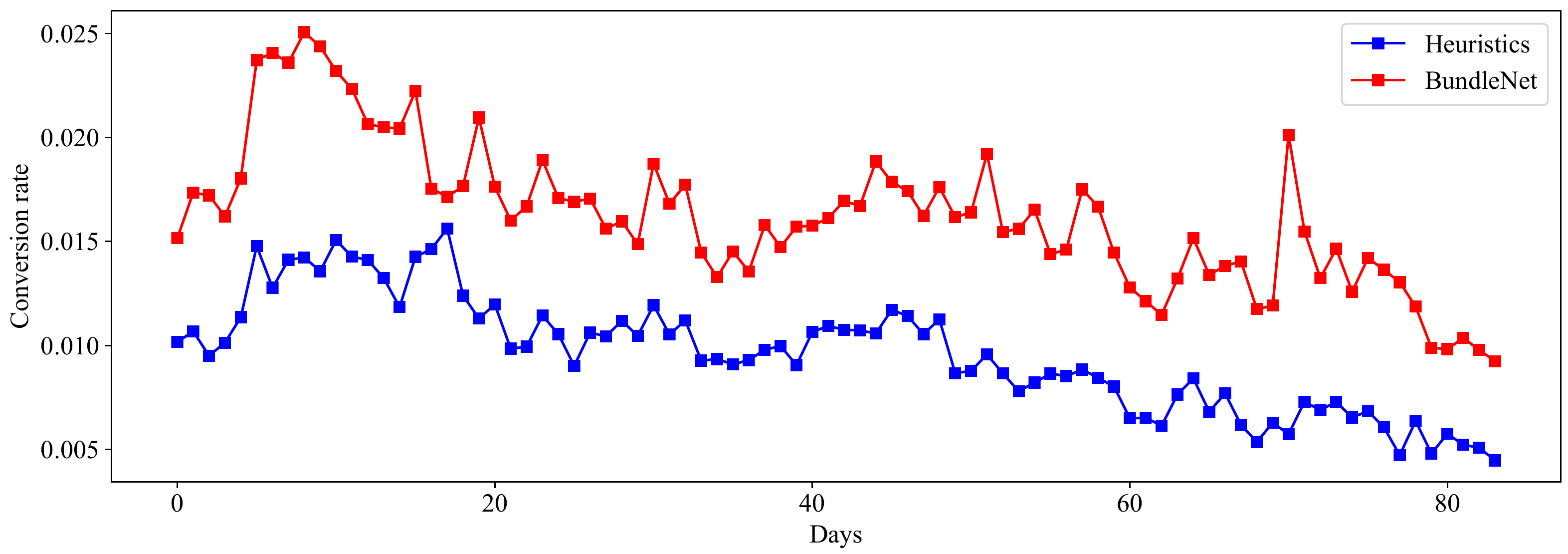}
  \label{fig:ctcvr}
  }
  \subfigure[Comparison of Conversion Rate of Bundles with Different Prices (from left to right arranged in ascending bundle prices).]{
  \includegraphics[width=0.48\textwidth]{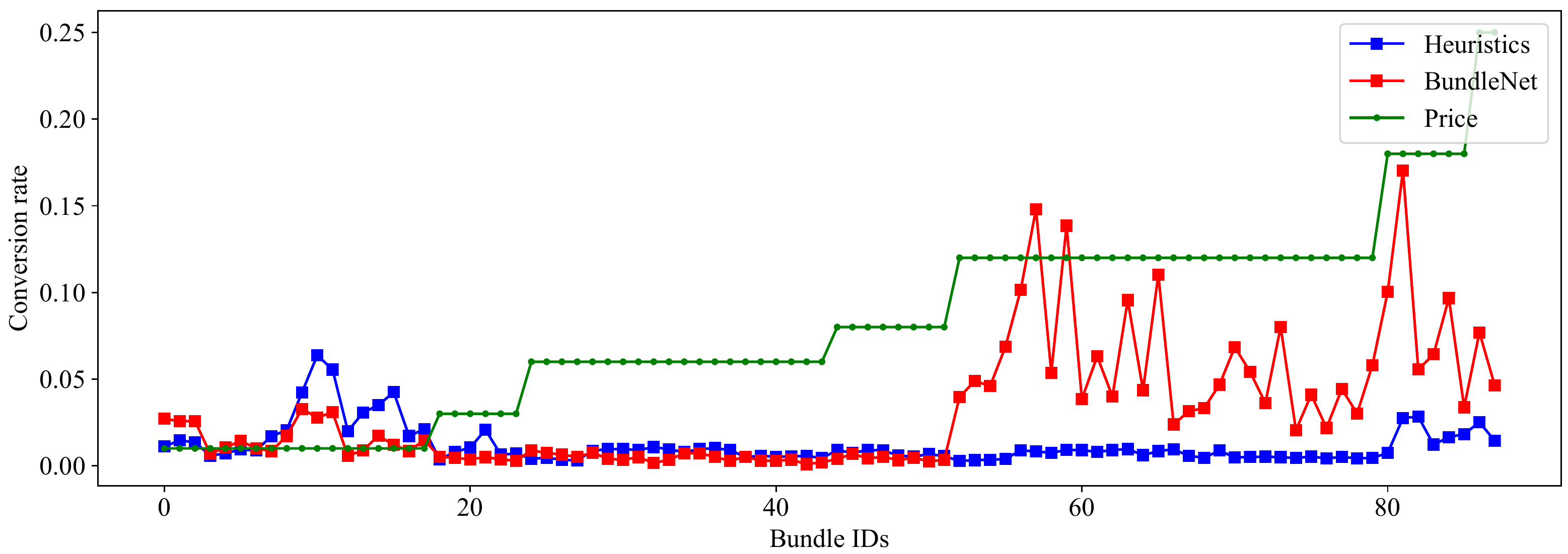}
  \label{fig:ctcvr_detail}
  }
  \caption{Online Performance of BundleNet on the Game \emph{Love is Justice}.} 
  \label{fig:online}
  \Description{Online Performance.}
\end{figure*} 

The proposed recommendation model has been deployed in production for more than one year in \emph{Love is Justice} developed by Netease Games. In this section, we briefly give some implementation details of the bundle recommendation pipeline in the online game, as shown in Figure \ref{fig:pipeline}.

\begin{itemize}
  \item Data Module. The data module is responsible for data storage and preprocessing tasks. The historical interaction data between users and items as well as bundles within a period of time is used to generate training data.
  \item Training Module. We train and update our recommendation model on a daily basis. Since retraining the model from scratch every time is computationally time-consuming, a better solution is to use the previously saved model as pre-training, and fine-tune the model on new data every day, which leads to faster convergence of model training.
  \item Serving Module. Once the model is trained and verified, we predict the preference scores, which are obtained by running a forward inference pass over the model, of all the bundles for all users. Then, the personalized bundles are ranked from the highest scores to the lowest, and the result is stored into the database for quick retrieval.
\end{itemize}

Figure~\ref{fig:ctcvr} gives the online performance within a period of nearly three months of the presented approach compared with a heuristic method in production, which is a combination of handcrafted recommendation rules. We can find that our proposed method always outperforms the heuristic method in online A/B testing. According to our analysis of purchase statistics, the launch of the model yields more than 60\% improvement on conversion rate (CVR) of bundles on average, and a relative improvement of more than 15\% in terms of gross merchandise volume (GMV).

For an in-depth analysis of the improvement, we calculate the conversion rate of most representative bundles with different prices separately. As shown in Figure~\ref{fig:ctcvr_detail} (the specific price values are properly processed on the scale for privacy considerations), we can find that the main reason for the improvement lies in the accurate recommendation of high-priced bundles. These bundles often contain more valuable items that are very attractive to players interested in them. Different from the lower-priced bundles which usually only contain common items, the high-priced bundles are highly personalized which leaves room for improvement. We also noticed that the purchase rate of low-priced bundles is higher than that of middle-priced bundles. We speculate that the types of items included in these bundles are not much different, but low-priced bundles are more appealing in price.

\section{Related Work} \label{sec:related}

In the field of recommendation, there have been several efforts to solve the problem of bundle recommendation. The \emph{List Recommendation Model} (LIRE) \cite{liu2014recommending} solves the recommendation problem of user-generated item lists based on a latent factor-based BPR model, which takes into consideration users’ previous interactions with both item lists and  individual items. \emph{Embedding Factorization Model} (EFM) \cite{cao2017embedding} is proposed to jointly model the user-item and user-list interactions, which combines two types of latent factor models: BPR \cite{rendle2009bpr} and word2vec \cite{mikolov2013distributed}. Also building upon the BPR model, \cite{pathak2017generating} trys to recommend existing bundles to users on the basis of their constituent items, as well as the more difficult task of generating new bundles that are personalized to a user via the bundle-level BPR model, which makes use of the parameters learned through the item-level BPR model. \emph{Deep Attentive Multi-Task} DAM \cite{chen2019matching} model designs a factorized attention network to aggregate the embeddings of items within a bundle to obtain the bundle’s representation, while jointly model user-bundle interactions and user-item interactions in a multi-task manner to alleviate the scarcity of user-bundle interactions. Some other related efforts include \cite{garfinkel2006design, sar2016beyond, beladev2016recommender, qi2016recommending, he2019hierarchical}. 

\section{Conclusion and future work} \label{sec:conclusion}

In this paper, we target at a practical but less explored recommendation problem named bundle recommendation. Different from the traditional item recommendation problem, it aims to recommend a \emph{bundle} (i.e., a combination of items) rather than the individual item to the target user. To tackle this specific recommendation problem instance in the context of the virtual mall in online games, we highlight the challenges and formalize it as a link prediction problem on a user-item-bundle tripartite graph, which is constructed from the historical interactions, and solve it within an end-to-end graph neural network framework. Extensive offline and online experiments demonstrate the effectiveness of the presented method. 

%%
%% The acknowledgments section is defined using the "acks" environment
%% (and NOT an unnumbered section). This ensures the proper
%% identification of the section in the article metadata, and the
%% consistent spelling of the heading.
% \begin{acks}
% To Do.
% \end{acks}

% \newpage

% \balance

%%
%% The next two lines define the bibliography style to be used, and
%% the bibliography file.
\bibliographystyle{ACM-Reference-Format}
\bibliography{main}

%%
%% If your work has an appendix, this is the place to put it.
% \appendix

\end{document}